\documentclass[a4paper,11pt]{article}
\usepackage{pos}
\usepackage{lineno}
\usepackage{wrapfig}
\usepackage{subcaption}

\title{Unraveling the Unexpected Seasonal Variation of Multi-Muon Events at the NO$\nu$A Near Detector}
\ShortTitle{Unraveling the Unexpected Seasonal Variation of Multi-Muon Events at the NO$\nu$A ND}

\author*[a]{Jordi Tuneu}
\author[a]{Eva Santos}
\author[a]{Peter Filip}

\affiliation[a]{Institute of Physics of the Czech Academy of Sciences,\\
  Na Slovance 1, Prague, Czech Republic}

\emailAdd{tuneu@fzu.cz}
\abstract{
The seasonal variation of single muons is a well-understood phenomenon, mainly driven by a positive correlation with temperature 
fluctuations in the atmospheric profiles. 
However, the rate of multi-muon events recorded by various experiments has shown an intriguing opposite seasonal modulation that 
remains unexplained by previous research.
For the first time, we quantitatively describe the phase and amplitudes of the seasonal variation for cosmic multi-muon events detected by the 
NO$\nu$A Near Detector. 
We can further explain the amplitude dependence for multi-muon events across various multiplicities. 
For our analysis, we use the general-purpose Monte Carlo code FLUKA-CERN 4.4.1, which provides a more realistic description of the detector, 
atmospheric profiles, and muon propagation underground. 
Finally, we compare our results with those obtained from the latest CORSIKA version 7.8010, utilizing the most up-to-date high-energy hadronic 
interaction models EPOS LHC-R and QGSJETIII-01. 
Our findings provide a fresh perspective on seasonal muon flux modulation and offer key constraints for cosmic-ray interaction models and underground detector studies.
}

\ConferenceLogo{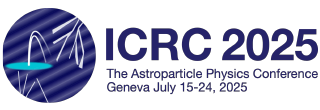}

\FullConference{39th International Cosmic Ray Conference (ICRC2025)\\
 15–24 July 2025\\
Geneva, Switzerland\\}
\begin{document}
\maketitle

\section{Introduction}
\vspace*{-0.3\baselineskip}
The seasonal variation of single muons, primarily driven by temperature-dependent effects in atmospheric density profiles, is a well-established phenomenon. 
As the atmosphere warms during the summer months, the probability of meson decay rather than further re-interaction increases due to the reduced air density in the upper atmosphere, leading to an enhanced flux of high-energy muons and the observation of a single-muon maximum in underground experiments. 
This behavior results in a positive correlation between muon rate and atmospheric temperature~\cite{Adamson2015, Acero2019}. 
MINOS cosmic data~\cite{Adamson2014} show that the total (single) muon flux increases in summer and decreases in winter by $\approx 1\%$. 
However, a puzzling and persistent discrepancy emerges when analyzing multi-muon events. 
Measurements from the MINOS and NO$\nu$A Near Detectors, both located approximately 100 meters underground at Fermilab, have revealed that the rate of multi-muon events follows an inverted seasonal variation. 
Specifically, the multi-muon event flux reaches a maximum in winter and a minimum in summer, with a modulation amplitude a factor of four larger than the one observed for single muons~\cite{Adamson2015, Acero2019, NOvA:2025nop}. 

Several hypotheses have been proposed to explain this unexpected anti-correlation~\cite{Adamson2015}. 
These include geometric effects linked to the seasonal shift in the altitude of the first interaction, differences in the production heights of multi- versus single-muon events, and the possible contribution of prompt dimuon decays from short-lived mesons. 
However, the most plausible explanation points to a competition between meson decay and interaction during the air shower development. 
The single-muon flux primarily results from the decay of leading hadrons, whereas multi-muons are typically generated by subsequent hadronic interactions that occur during later stages of shower development. 
In summer, the warmer atmosphere increases the probability that secondary mesons will decay before interacting, which in turn decreases the production of pions and kaons, which would potentially lead to the generation of additional muons.
This seasonal suppression of hadronic branching in summer naturally leads to a reduced multi-muon rate, and conversely to a winter maximum. 

In this work, we revisit the multi-muon seasonal variation puzzle using a new simulation approach based on the general-purpose Monte Carlo code FLUKA-CERN 4.4.1~\cite{Flukacr}. As the FLUKA~\cite{Flukacr, Ballarini:2024isa} codes allow a more detailed and realistic treatment of the atmospheric density profile, 
particle interactions, and transport in multiple media, as well as customized detector geometry, they provide a more accurate representation of the phenomenon.
For the first time, we quantitatively describe the phase and amplitude of the seasonal variation in multi-muon events detected by the NO$\nu$A Near Detector (ND). 
We further examine the dependence of the modulation amplitude on muon multiplicity and cosmic-ray composition, revealing a strong role played by the contribution of heavier nuclei to the observed seasonal effect. 
Finally, we compare our results with those obtained using the latest CORSIKA~7.8010~\cite{corsika} version, with the EPOS LHC-R~\cite{EPOS}, and QGSJETIII-01~\cite{QGSJET} high-energy hadronic interaction models, and FLUKA~2024.1.3~\cite{Ballarini:2024isa} as the low-energy hadronic interaction model. 

\section{FLUKA simulation layout}\label{model}
\begin{wrapfigure}{r}{0.5\textwidth}
  \begin{center}
    \includegraphics[width=0.5\textwidth]{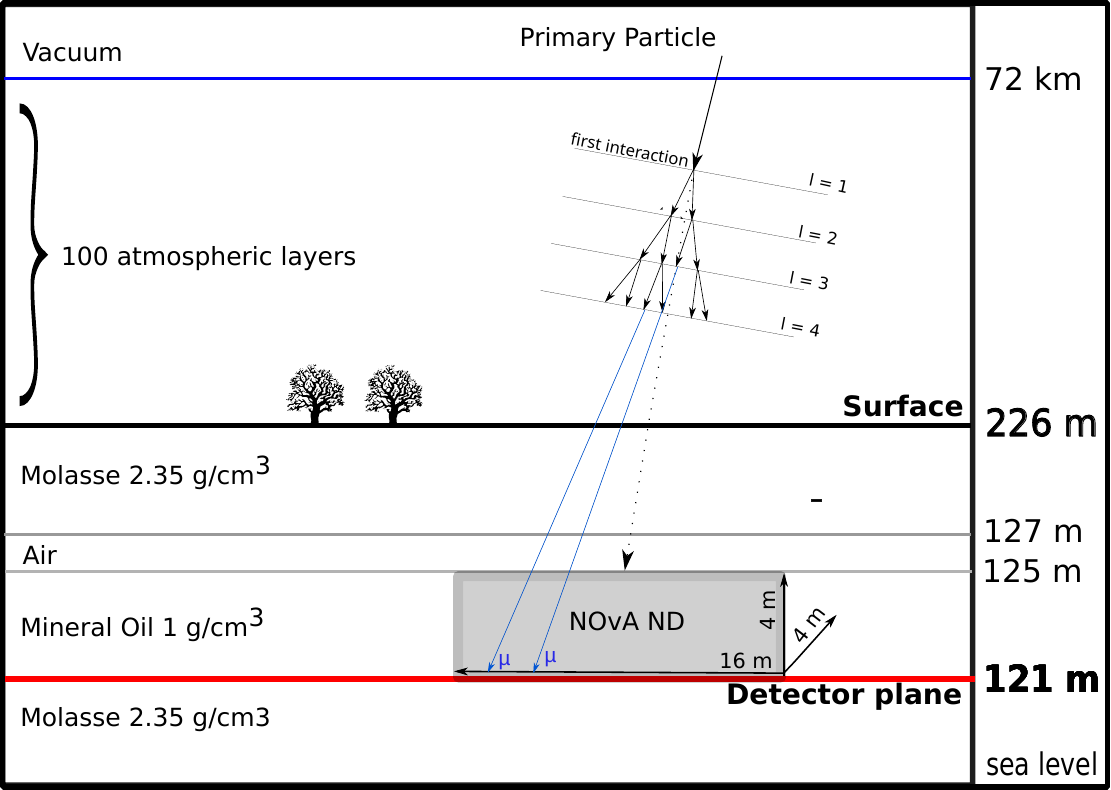}
  \end{center}
  \caption{Schematic layout of the FLUKA geometry used to simulate muon propagation to the NO$\nu$A Near Detector (not to scale). The simulation includes a 99\,m molasse overburden, a 2\,m air-filled detector cavern, a 4\,m mineral oil layer, and additional molasse extending to sea level.}
  \label{fig:geometry}
\end{wrapfigure}

We utilize the FLUKA-CERN version 4.4.1~\cite{Flukacr} Monte Carlo code to simulate the seasonal 
variations in multi-muon events at the NO$\nu$A Near Detector, accounting for the specific atmospheric
and underground conditions at the Fermilab site during the data-taking period. 
This setup, whose schematic sketch is shown in Figure~\ref{fig:geometry}, provides a more realistic
description of the atmospheric profile density and mass overburden for our simulation of the multi-muon
event rates at the NO$\nu$A Near Detector, all within a single Monte Carlo code.

The atmospheric profile is modeled using a 100-layered stratified structure, ranging from an altitude of 72\,km down to ground level at 226\,m above sea level. 
For each layer, the air density is calculated from the European Centre for Medium-Range Weather Forecasts (ECMWF) ERA5 database~\cite{ECMWF}, utilizing the geopotential height and temperature measured at 37 isobaric levels at 6-hour intervals.
We extracted the atmospheric temperature and geopotential height data for January, July, and the whole year of 2016, and computed the monthly and yearly mean density profiles. 

These profiles were then adjusted to align with the 100-layer atmospheric model employed in FLUKA, resulting in season-specific density profiles for the simulation. 
To simulate the propagation of muons underground, we implemented a layered detector geometry that approximates the setup of the NO$\nu$A Near Detector. 
The first underground layer consists of 99\,m of molasse, extending from the surface to the detector cavern at an elevation of 127\,m above sea level, followed by a 2-meter-high air-filled cavity where the detector is located, and an additional 4-meter layer of mineral oil below the detector. 
A final molasse layer extends from the bottom of the oil layer down to sea level, absorbing the remaining particle flux. 
The entire simulation volume is embedded in a cylindrical geometry with a uniform magnetic field with components set to \( B_x = 19.3\,\mu\mathrm{T} \), \( B_y = 0 \), and \( B_z = 49.7\,\mu\mathrm{T} \), derived from the IGRF2020 model for the Fermilab site in 2017. 

\subsection{FLUKA cosmic-ray shower library}\label{library}
Our cosmic ray library comprises a four-component mass composition: namely, proton, helium, nitrogen (representing the CNO group), and iron. 
Cosmic rays were simulated following a power-law spectrum with spectral index $\gamma = 2.7$, ranging from 0.5\,TeV per nucleon up to a maximum energy of 
1\,PeV, the highest energy allowed for FLUKA-CERN simulations. 
The zenith angle distribution ranges from $20^{\circ}$ to $60^{\circ}$, following a uniform distribution in $\sin^{2}\theta$, compatible with the one
observed by a flat detector. 
The azimuthal angular distribution is uniform between $0^{\circ}$ and $360^{\circ}$. 
For our studies, the elemental groups are injected with relative weights chosen to reflect the cosmic-ray abundances measured at an energy of 1\,TeV per 
nucleon, according to~\cite{AMS2015, CREAM2017}, where the mass composition is dominated by light nuclei.

Each cosmic ray is injected along a horizontal ring located at an altitude of 72\,km, with a surface area of $20\,\mathrm{km} \times 20\,\mathrm{km}$, 
which projects to approximately $2\,\mathrm{km} \times 2\,\mathrm{km}$ at the detector plane (121 m a.s.l.), ensuring uniform exposure over the full 
zenith-angle range. 
The position of the primary particle is chosen so the shower axis intersects the surface at $(x, y, z) = (x^{\prime}, y^{\prime}, 125)$, and the 
detector plane at coordinates  $(x = y = 0)$, yielding consistent geometry across simulations. 
Each shower is propagated through the layered atmosphere and underground geometry described above (see also Figure~\ref{fig:geometry}), 
and muons are tracked until they either reach the NO$\nu$A Near Detector or fall below the energy threshold required to penetrate the mass
overburden.

For each seasonal profile, we simulate a total of 900,000 proton-induced showers, with additional primary species (He, N, and Fe) generated according to the
relative abundance ratios from~\cite{AMS2015, CREAM2017}. 
In order to improve the statistical stability in the estimation of multi-muon observables, a larger number of proton-induced showers was needed, as very few
of them were found to generate dimuon events at the lowest energies. 
Moreover, particularly at high multiplicities where event rates are low, we implement a bootstrap resampling procedure. 
For a given observable and angular-multiplicity bin, if $N_\text{sim}$ independent showers are available and $N_\text{use} \leq N_\text{sim}$ are required 
per evaluation, we generate $B$ bootstrap samples by selecting, without replacement, $N_\text{use}$ unique showers from the pool of $N_\text{sim}$. 
This process is repeated independently for each sample. 
The observable is then computed for each bootstrap realization, and the final result is obtained by averaging over the $B$ samples. 
This procedure enhances statistical robustness without requiring an increase in the number of full shower simulations.
This shower library provides the basis for all subsequent analyses, including seasonal flux comparisons and multiplicity-dependent studies of multi-muon events. 
The number of showers per elemental group is summarized in Table~\ref{tab:shower_summary}.
\begin{table}[h]
\centering
\caption{Summary of simulated showers per primary species. 
Here, $N_{\text{sim}}$ refers to the total number of showers available from the full simulation library, 
$N_{\text{use}}$ is the number of unique showers used per bootstrap sample in the analysis, 
and the bootstrap factor denotes the number of resamplings performed. 
This procedure improves statistical stability without requiring an increase in the number of full simulations.}
\label{tab:shower_summary}
\begin{tabular}{lccc}
\hline
\textbf{Primary} & $N_{\text{sim}}$ & $N_{\text{use}}$ & Bootstrap factor \\
\hline
p   & 1,000,000 & 900,000 & 1 \\
He  & 125,000   & 90,000  & 10 \\
N   & 8,000     & 6,000   & 50 \\
Fe  & 2,000     & 1,200   & 200 \\
\hline
\end{tabular}
\end{table}

\vspace*{-0.3\baselineskip}

\section{FLUKA results}
In this section, we present our results on the seasonal modulation of multi-muon events at the NO$\nu$A Near Detector, using the FLUKA-CERN~4.4.1 
Monte Carlo code, with the detector layout and shower library described in Section~\ref{model}. 
Our objective is to quantitatively compare the predicted seasonal variation in multi-muon event rates with that observed experimentally, with a specific focus on the amplitude and multiplicity dependence of this effect, reproducing the conditions for the January and July 2016 data-taking. 
%First, we assess the overall muon multiplicity distribution in both winter and summer atmospheric conditions for the full shower library described in Section 2.1. 
Our results are compared with those from~\cite{Acero2019}, which are described by the following 
seasonal variation:
\begin{equation}
f(t) = V_0 + V \cos(\omega t + \phi), \label{eq:fit}
\end{equation}
where $V_0 = 0 \pm 0.1\%$, $V = 4.1 \pm 0.2\%$, and $\phi = -0.43 \pm 0.05$, corresponding to a maximum 
near January 25 and a minimum near July 26. 

\subsection{Multi-Muons: Seasonal Variation}\label{fluka_results}
To increase statistical precision and fully explore the distance range of muons arriving in coincidence
at the detector plane, contemplating several configurations, we employ a grid of virtual detector volumes
with dimensions $16 \times 4$\,m$^2$, matching the size of the NO$\nu$A Near Detector, tiled over a 
$2\,\mathrm{km} \times 2\,\mathrm{km}$ region at 125\,m depth. 
This method enables the evaluation of each simulated shower at multiple transverse positions, effectively
sampling a wide range of muon multiplicities and lateral distributions. 
By scanning all spatial configurations that would lead to multiple muons entering a NO$\nu$A-sized
detector, we account for the statistical spread of shower core locations and muon lateral density
profiles without relying on a fixed detector position.

\begin{figure}[htpb]
\centering
\includegraphics[width=0.5\textwidth]{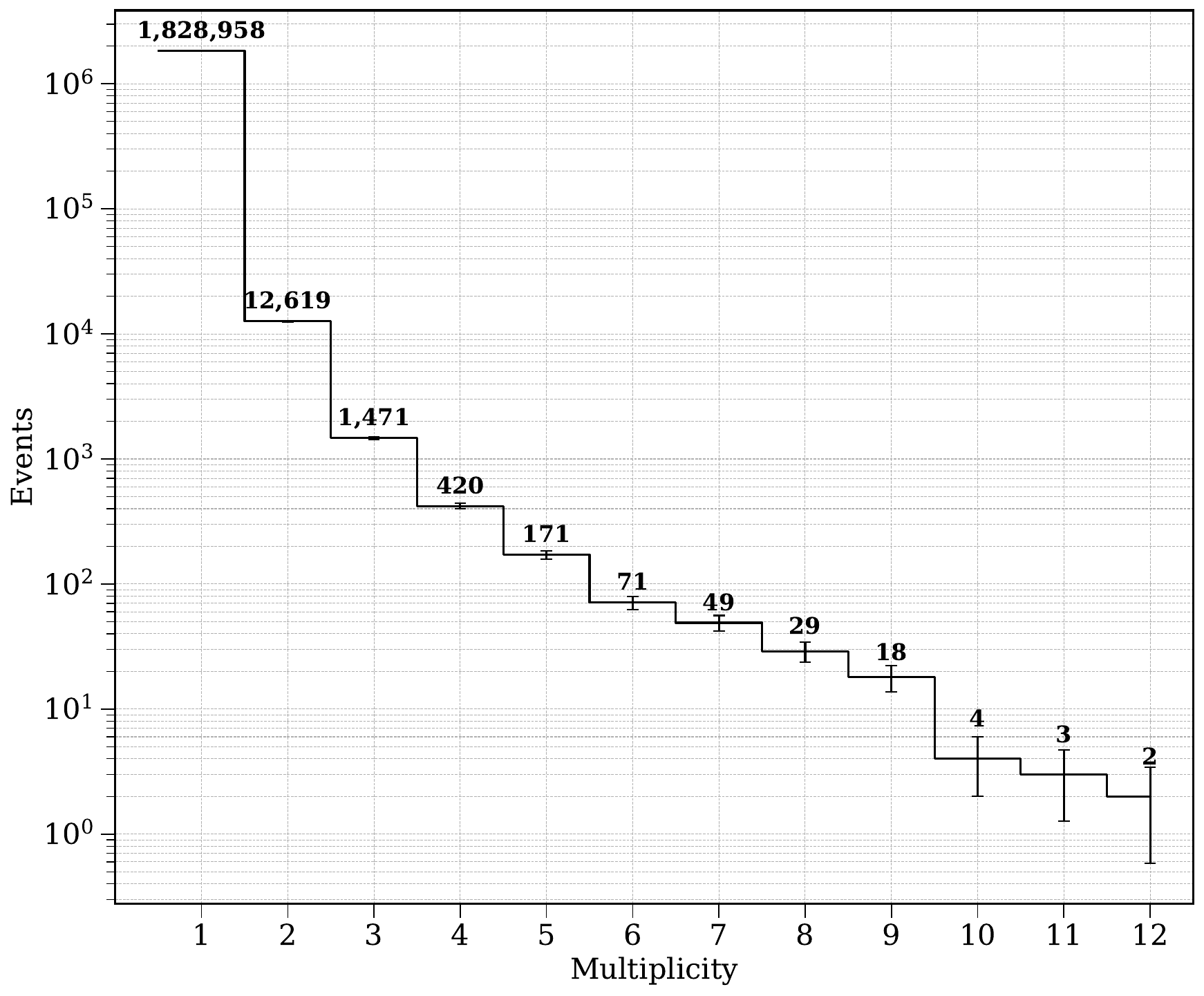} 
\caption{
Simulated multiplicity distribution of muon events at the No$\nu$A Near Detector underground depth for the winter and summer atmospheric profiles. The distribution includes muons induced by primary proton, helium, nitrogen, and iron nuclei, weighted according to the relative composition described in subsection~\ref{library}. The plotted multiplicities correspond to the number of muons simultaneously intersecting a virtual detector plane of $16 \times 4$\,m$^2$ at 99\,m depth. The vertical axis is shown on a logarithmic scale to highlight the range across multiplicities.
}
\label{fig:multiplicity_distribution}
\end{figure}

In Figure~\ref{fig:multiplicity_distribution}, we show the simulated multiplicity distribution of muon 
events at the depth of the No$\nu$A Near Detector for both winter (January) and summer (July) 
atmospheric profiles. 

In Figure~\ref{fig:multiplicity_distribution}, the events are aggregated for all primary types (p, He, N, Fe), according to the relative weights defined
in subsection~\ref{library}. 
Heavier nuclei, such as iron or nitrogen, tend to produce air showers with a larger number of muons, 
leading to a larger multiplicity spectrum. 
As expected, lower multiplicities are dominant, while higher-multiplicity events become increasingly 
rare due to the steeply falling cosmic-ray particle spectrum and the lateral spread of muons in air showers.

\begin{figure}[htpb
]
\centering
\includegraphics[width=1.0\textwidth]{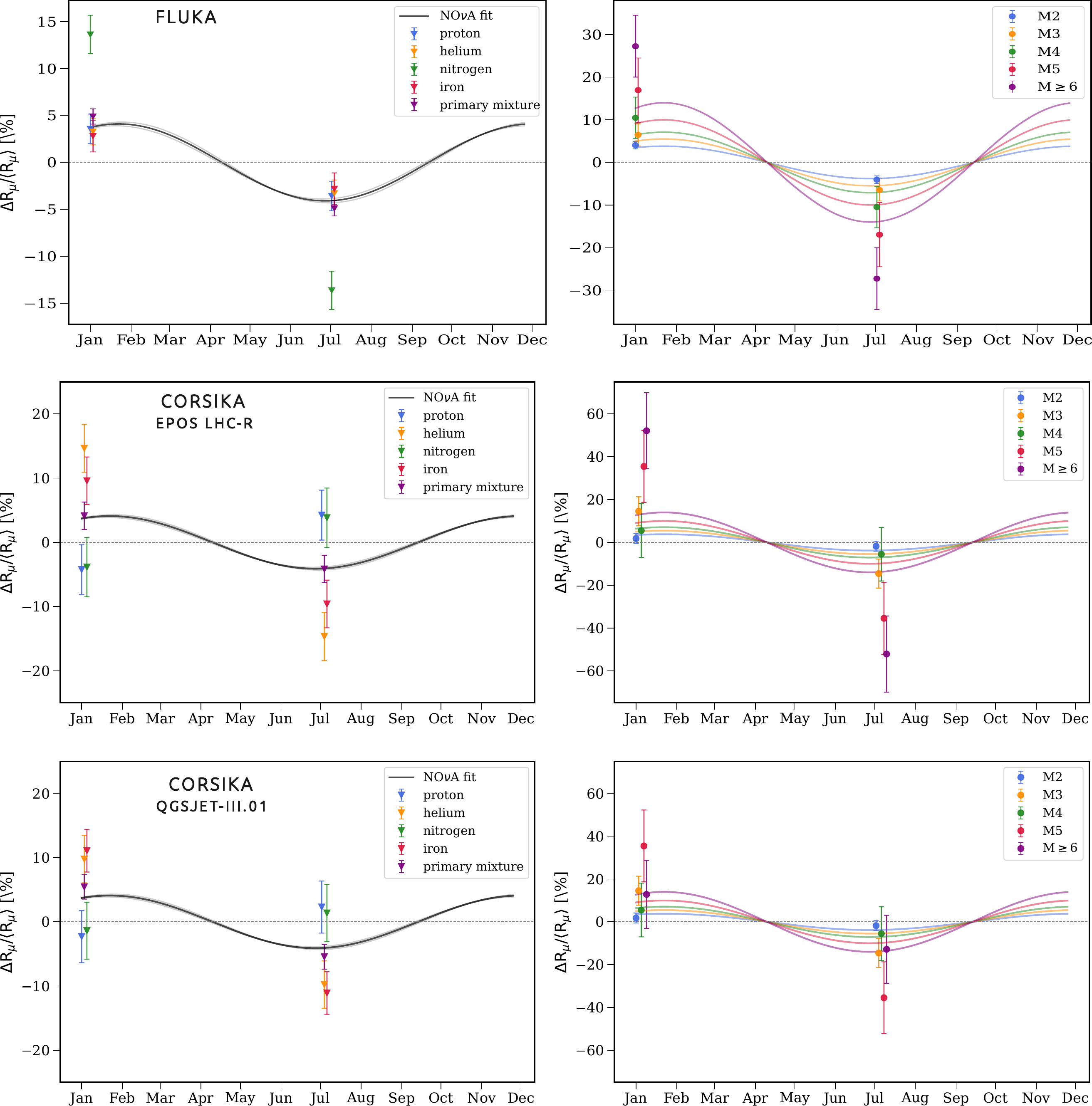}
\caption{
Simulated seasonal variation of the multi-muon rate at the NO$\nu$A Near Detector depth obtained by, from top to bottom, FLUKA, CORSIKA EPOS LHC-R, CORSIKA QGSJETIII.01. From left to right: Relative modulation for different primary species. Modulation across muon multiplicity classes (from m$=$2 to m$\geq 6$. All curves represent the best-fit calculated by NO$\nu$A~\cite{Acero2019}.
}
\label{fig:results}
\end{figure}

The seasonal modulation of the multi-muon rate in our simulations shows variations that depend on both
the primary composition and the muon multiplicity. 
In Figure~\ref{fig:results} (first row left), the relative seasonal variation as a function of the two
months (January and July) for showers initiated by individual primary species is shown. 
While proton, helium, and iron show amplitudes consistent with the No$\nu$A Near Detector data within
statistical uncertainties, the modulation observed for nitrogen is significantly higher, with an
amplitude of $13.63 \pm 2.04\%$. 
The origin of this behavior is not yet understood, and it is under investigation. 
The combined result from all simulated species, weighted according to the composition model described 
in subsection~\ref{library}, agrees well with the NO$\nu$A Near Detector data~\cite{Acero2019} described
by Equation~\ref{eq:fit}, both in phase and amplitude.

The multiplicity dependence of the modulation is shown in Figure~\ref{fig:results} (first row, right). 
As the number of muons in an event increases, so does the amplitude of the seasonal variation. 
This trend is consistent with the No$\nu$A Near Detector observations~\cite{Acero2019}, and it is expected
from the increased role of high-energy and heavy nuclei in producing higher-multiplicity events. 
At the same time, the statistical uncertainty increases with multiplicity due to the reduced number of
high-multiplicity events available in both simulation and data. 
In simulations, the contributions from nitrogen and iron become increasingly relevant for $M \geq 4$, 
as lighter primaries, such as proton and helium, rarely produce more than 3–4 muons in a single 
detector-sized volume. 
A realistic cosmic-ray composition is therefore crucial to reproduce the whole structure of the observed
modulation across multiplicities.

\section{CORSIKA simulation layout and results}
We further evaluated the seasonal modulation of multi-muon events using the most up-to-date 
CORSIKA~v7.8010~\cite{corsika} with the hadronic interaction models EPOS LHC-R~\cite{EPOS} 
and QGSJETIII-01~\cite{QGSJET}. 
The simulations follow the same primary composition, energy spectrum, and multiple detector tiling adopted in the FLUKA analysis. 
In this case, we adopted a custom 5-layered atmospheric profile, which resulted from the best fit to 
the averaged data extracted from the European Centre for Medium-Range Weather Forecasts (ECMWF) ERA5
database~\cite{ECMWF}. 
However, these data only offer measurements up to a height of about 40\,km, leaving a small lever arm to 
establish a good fit for the 4$^{\mathrm{th}}$ layers, whose range typically spanned from about 
20 to 100\,km. 
It turned out that the estimated atmospheric depth at 100\,km is about $10\,\mathrm{g\,cm^{-2}}$, a 
very unrealistic value at such high altitudes. 
Consequently, we had to define an extra custom 5$^{\mathrm{th}}$ layer, so that the atmospheric depth 
would naturally reach $0\,\mathrm{g\,cm^{-2}}$ at about 112\,km, the pre-defined top of the atmosphere in 
CORSIKA, meaning that the CORSIKA atmospheric profiles are approximately $10\,\mathrm{g\,cm^{-2}}$ larger
than those in FLUKA. 
Another difference between the two models is the fact that CORSIKA only allows for the development of
extensive air showers in air. 
Therefore, in order to match the FLUKA results, we applied an energy cut for the muons at the ground 
level defined as $E^{\mathrm{det}} = E^{\mathrm{surf}} / \cos \theta$, where $E^{\mathrm{det}}$ is the estimated 
energy of the muons at the detector level, $E^{\mathrm{surf}}$ is the energy of the muon at the ground level, and 
$\theta$ is the zenith angle of the incident cosmic ray. 
All muons with a vertical energy at the ground below 50\,GeV were excluded from the analysis.

The results for EPOS LHC-R and QGSJETIII-01 are shown in Figure~\ref{fig:results}, middle and bottom
rows, respectively. 
From inspection of Figure~\ref{fig:results}, we see that both FLUKA and the two high-energy hadronic
interaction models  EPOS LHC-R and QGSJETIII.01, from the CORSIKA simulations, are able to reproduce the
overall seasonal variation observed by NO$\nu$A Near Detector. 
However, in CORSIKA simulations, the behavior of individual primaries differs slightly between the two 
models and also deviates from that of the FLUKA simulations. 
In both cases, proton- and nitrogen-initiated showers show their maxima in July, while helium and iron
primaries peak in January, in line with the expected maximum of the observed seasonal variation. 
Despite these differences at the level of single components, the composition-weighted mixture lies 
close to the NO$\nu$A results, both in phase and in amplitude. 
The multiplicity dependence is also qualitatively reproduced, with the modulation amplitude increasing with the number of muons. 
The limited statistics at high multiplicity prevents firm conclusions, but the trend is clearly visible 
and consistent with the data. The level of agreement is encouraging, although it does not reach the 
consistency achieved with FLUKA.

It should be emphasized that multiplicity-dependent results are intrinsically more volatile than 
single-muon modulations, both due to the limited event statistics at higher multiplicities and to the 
strong variability associated with sampling only a restricted volume of the detector. 
These constraints need to be taken into account when comparing different models and when assessing the level of agreement with data.

\section{Conclusions}
For the first time, using the general-purpose FLUKA-CERN 4.4.1 Monte Carlo code, we successfully reproduced both the phase amplitude and 
overall modulation of the multi-muon seasonal observations reported first by the MINOS Near Detector~\cite{Adamson2015}, and confirmed by 
the NO$\nu$A Near Detector~\cite{Acero2019, NOvA:2025nop}, and also the phase and amplitudes for several muon multiplicities. 
In a subsequent phase, we also evaluated the performance of CORSIKA 7.8010, utilizing EPOS LHC-R and QGSJETIII-01 high-energy hadronic 
interaction models, combined with FLUKA 2024.1.3 to treat hadronic interactions below 30/80\,GeV (EPOS LHC-R/QGSJETIII.01). 
We found that, while CORSIKA is also able to reproduce the overall phase amplitude and multi-muon multiplicity, there are differences for 
the different elemental species, even when compared with FLUKA predictions. 
These results may stem from both the different physics and the high-to-low energy transition of the two models. 
Further studies with Sibyll 2.3e~\cite{sibyll} and DPMJET~\cite{DPMJET}, the latter combined with UrQMD~\cite{urqmd} in CORSIKA for an energy 
transition between models of 50 GeV, are underway. 
Our simulations do not reproduce the reported 10\% amplitude for proton-induced showers claimed in~\cite{NOvA:2025nop}.

\section*{Acknowledgments}
This work is funded by Czech Science Foundation under the project \mbox{GA\v{C}R 21-02226M}, and by the Czech Ministry of Education and Youth and Sport, within the projects Fermilab-CZ LM2023061, and CZ.02.01.01/00/22\_008/0004632.

\end{document}